\documentclass[journal]{IEEEtran}
\ifCLASSINFOpdf
\else
\fi
\usepackage{graphicx}  
\usepackage{bm}        
\usepackage{amssymb}   
\usepackage{amsmath}
\usepackage{float}
\usepackage{color}
\usepackage{physics}
\usepackage{xparse,xcolor}
\usepackage{array}

\begin{document}
%
\title{Free-space point-to-multiplepoint optical frequency transfer with lens assisted integrated beam steering}
%
%
%

\author{Liang Hu,~\IEEEmembership{Member,~IEEE,} Ruimin Xue, Xianyi Cao, Jiao Liu, Kan Wu,~\IEEEmembership{Member,~IEEE,} Guiling Wu,~\IEEEmembership{Member,~IEEE} and Jianping Chen
\thanks{Manuscript received xxx xxx, xxx; revised xxx xxx, xxx. This work was supported by the National Natural Science Foundation of China (NSFC) (62120106010, 61905143), the Zhejiang provincial Key Research and Development Program of China (2022C01156) and the National Science Foundation of Shanghai (22ZR1430200). (Corresponding author: Jianping Chen)}
\thanks{Ruimin Xue, Xianyi Cao, Jiao Liu, and Kan Wu are with  the State Key Laboratory of Advanced Optical Communication Systems and  Networks, Department of Electronic Engineering, Shanghai Jiao Tong University,  Shanghai 200240, China (e-mail: rmxue96@sjtu.edu.cn; mr.cao@sjtu.edu.cn; liujiao@sjtu.edu.cn; kanwu@sjtu.edu.cn)} 
\thanks{Liang Hu, Guiling Wu, and Jianping Chen are with  the State Key Laboratory of Advanced Optical Communication Systems and  Networks, Department of Electronic Engineering, Shanghai Jiao Tong University,  Shanghai 200240, China, and also with the SJTU-Pinghu Institute of Intelligent Optoelectronics, Jiaxing Time-Transfer Photoelectric Co., Ltd., Pinghu 314200, China, China (e-mail: liang.hu@sjtu.edu.cn;  wuguiling@sjtu.edu.cn; jpchen62@sjtu.edu.cn)} 
}

%
%

\markboth{preprint, 2022}%
{Shell \MakeLowercase{\textit{et al.}}: Bare Demo of IEEEtran.cls for IEEE Journals}
%



\maketitle

\begin{abstract}
We report on the realization of  high-performance silica integrated two-dimensional lens assisted beam-steering (LABS) arrays along with the first-of-their-kind point-to-multiplepoint optical frequency transfer. {The LABS equips with  $N$ antennas} and has the capability to produce arbitrary number of output beams with different output angles with the simple control complexity.  We demonstrate that the LABS has 16 scanning angles, which can support {the access capability for the maximum of simultaneous 16 user nodes.}  The coaxial configuration for transmitting  and receiving the light  as a monolithic transceiver allows us to reduce the out-of-loop phase noise significantly.  Finally, the LABS-based non-blocking point-to-multiplepoint in-door free-space optical frequency transfer links with 24 m and 50 m free-space links are shown.  After being compensated for the free-space link up to 50 m, the fractional frequency instability of $4.5\times10^{-17}$ and $7.7\times10^{-20}$ at the averaging time of 1 s and 20,000 s, respectively, can be achieved. The present work proves the potential application of the 2D LABS in free-space optical time-frequency transfer and provides a guidance for developing a chip-scale optical time-frequency transfer system. 
\end{abstract}


%
\IEEEpeerreviewmaketitle

\section{Introduction}

\IEEEPARstart{U}{nprecedented} stability and uncertainty of optical neutral  atom and ion clocks can have the capability to accelerate high-precision measurements in applied  and fundamental  sciences \cite{grotti2018geodesy, Riehle:2017aa, delva2017test, lisdat2016clock, takamoto2020test}. However, these optical clocks have typically limited in size, weight, power and cost (SWaP-C) and are usually only available in national metrology institutes. To solve this dilemma situation, time-frequency transfer has widely attracted research attentions. Among different time-frequency transfer methods, optical fibers have been recognized as one of the most important host materials for high-performance time-frequency transfer \cite{ma1994delivering, predehl2012920, Calonico2014, cantin2021accurate, xu2019two, hu2020cancelling, fujieda2008ultrastable, sliwczynski2012frequency}. As fiber links are not always available, such as the time-frequency transfer or comparisons between optical atomic clocks in air channels, an alternative method for comparing or transferring atomic clocks over free-space links has to be investigated. Prerequisitely, this method would have the requirements in terms of the residual fractional instabilities and inaccuracies, which need to be  better than those of {the stability and accuracy of the clocks themselves in the presence of atmospheric turbulence. The capability to realize remote clock comparison or transfer over the free-space channel \cite{giorgetta2013optical, dix2021point, kang2019free, bergeron2019femtosecond, gozzard2018stabilized, sinclair2018comparing, deschenes2016synchronization} will promote the development of the fundamental physics such as  precise measurement of the variability of fundamental constants \cite{godun2014frequency},  gravitational waves \cite{delva2018gravitational, hu2017atom}, searches for dark matter \cite{derevianko2014hunting}, gravitational redshift \cite{takamoto2020test},  and just to name a few.

To develop a wireless connected  clock network, it is imperative to develop an appealing  method to distribute an ultrastable source to multiple user nodes. In optical fiber networks based optical frequency transfer, passive optical devices such as optical splitters could be used for achieving multiple-node optical frequency transfer over bus \cite{grosche2014eavesdropping, hu2020allpassive} and branching fiber networks \cite{schediwy2013high, hu2020multi, xue2021branching} { with active phase noise cancellation (ANC) \cite{grosche2014eavesdropping, schediwy2013high} or passive phase noise cancellation (PNC) \cite{hu2020allpassive, hu2020multi, xue2021branching}.} {Principally,  several point-to-point optical frequency transfer systems could be used to form a point-to-multiplepoint optical frequency transfer system, which will significantly increase the complexity of the system at the local site. Cascaded multiple systems can also provide the multiple-access capability such as comb-based two-way optical time-frequency transfer across a free-space channel by cascading two sets of free-space transceivers \cite{bodine2020optical}.} Moreover, free-space multiple-access radio frequency transfer by extracting the forward and backward signals has been demonstrated in \cite{hou2018free}. However, the access place's performance strongly depends on the main fiber link. {At the same time,} it is not convenient as the access place has to be on the free-space path \cite{hou2018free} and this is not always available or possible. One suspended question is how to distribute an ultrastable optical reference to multiple-user in a cost-effective and robust way.

 To date, all previous demonstrators of free-space optical frequency transfer  systems were based on discrete components \cite{giorgetta2013optical, dix2021point, kang2019free, swann2017low} and many active optical terminals have been proposed and experimentally demonstrated, such as mechanical galvanometers \cite{giorgetta2013optical, dix2021point, kang2019free} or {fast steering mirrors controlled by piezo actuators \cite{holmstrom2014mems, briles2010simple}}. However, current optical terminals use discrete components which limits deployability in  SWaP-C critical applications. Moreover, mechanical galvanometers pose several great challenges, consisting of limited beam steering precision, low stability and repeatability, and slow steering speed. Additionally, it is also great challenge by integrating mechanical galvanometers into other {components} for developing integrated optical transceivers because of their nonplanar structure. {Although the  fast steering mirrors controlled by piezo actuator can have the capability to mitigate the challenges posed by the mechanical galvanometer, it still has the limitations in  the SWaP-C critical applications.} In order to overcome these limitations, solid optical phased arrays (OPAs) for the beam steering applications has been widely proposed and demonstrated \cite{hutchison2016high, yaacobi2014integrated, xu2018wide, xu2019aliasing, komljenovic2017sparse, sun2013large, chung2017monolithically, poulton2019long, poulton2017coherent, phare2018silicon, poulton2017large, shin2020chip}.  In the previous pioneering work, the OPAs have attracted lots of research interest  in the application of  autonomous driving \cite{chen2017multi}, sensing \cite{lefsky2002lidar}, and so on.  By properly adjusting the relative phase delay between adjacent OPA's array elements,  usable beam steering with wide field of view \cite{yaacobi2014integrated, xu2019aliasing}, low divergence angle \cite{hutchison2016high}, and high side-mode suppression \cite{komljenovic2017sparse} can be achieved.  Currently, despite great progress, such as  scaling the OPA to a thousand elements, the high power consumption for the OPAs hinders their application because it requires to tune a large number of phase shifters. Taking an example, an OPA can consume tens of watts for the case which includes a thousand elements \cite{hutchison2016high, chung2017monolithically} if there is not  a particular technology employed to reduce the power consumption for each phase shifter \cite{miller2020large, chung2019low}.   Promisingly, a new beam-steering technique, namely lens assisted beam steering (LABS), based on an on-chip switch network and a lens \cite{inoue2019demonstration, li2019lens, ito2020wide, chang20212d} has attracted research interest.  The input light is distributed to the free-space via emitters by an integrated switch network \cite{inoue2019demonstration, li2019lens, ito2020wide, chang20212d}. Afterwards, the output beam from the emitter is collimated and deflected by the lens. The deflected angle is determined by the relative positions between the emitters and lens. Fortunately, the divergence of the beam based on the LABS can be easily tuned by changing the lens with different focal length \cite{inoue2019demonstration, li2019lens, ito2020wide, chang20212d}.  
 

Although the optical beam steering has demonstrated its promising characteristics, most of the optical beam steering techniques have adopted the separate transmitting and receiving circuits, enabling the increase in the sensitivity of the receiver. Different from the communication applications, optical time-frequency transfer is extremely sensitive to the outside path, which has been taken as optical interferometer noise \cite{williams2008high, stefani2015tackling}. Consequently, it is important to shorten the outside path as short as possible. The optical interferometer noise has previously observed in fiber and free-space based time-frequency transfer systems \cite{stefani2015tackling, xu2018studying}. Several methods have been proposed to overcome the interferometer noise.  By passive and active temperature stabilization  interferometers \cite{stefani2015tackling, xu2018studying, hu2021performance}, the temperature {sensitivity}  can be effectively suppressed by a factor of 7, reaching to 1 $f$s/K. Alternative promising method is to significantly shorten the out-of-loop path by integrating the optical paths and components into a single photonic integration chip as recently demonstrated in  photonic {planer} lightwave circuits \cite{akatsuka2020optical}. We recently demonstrated the interferometer noise floor of  $2.3\times10^{-17}$ at 1 s and $6.3\times10^{-21}$ at 40,000 s by integrating the outside paths into silicon circuits \cite{hu2021silicon}, which is almost one order of magnitude better than our temperature-stabilized fiber interferometer \cite{hu2021performance}. All the after-mentioned methods need complicated temperature stabilization methods or dedicated photonic integration technique  to reduce the interferometer noise. As we discussed in \cite{xue2021branching}, the simplest and most robust way to reduce the interferometer noise is to include the outside paths or parts into the in-loop. By optimizing the system in this way, the phase noise coming from the outside-loop can be effectively suppressed by adopting the active or passive fiber phase noise cancellation technique \cite{ma1994delivering, hu2020cancelling}. Therefore, the monolithic optical transceiver by sharing the transmitting and receiving path will be ideal solution to overcome the after-mentioned challenges.

 \begin{figure*}[htbp]
\centering
\includegraphics[width=0.97\linewidth]{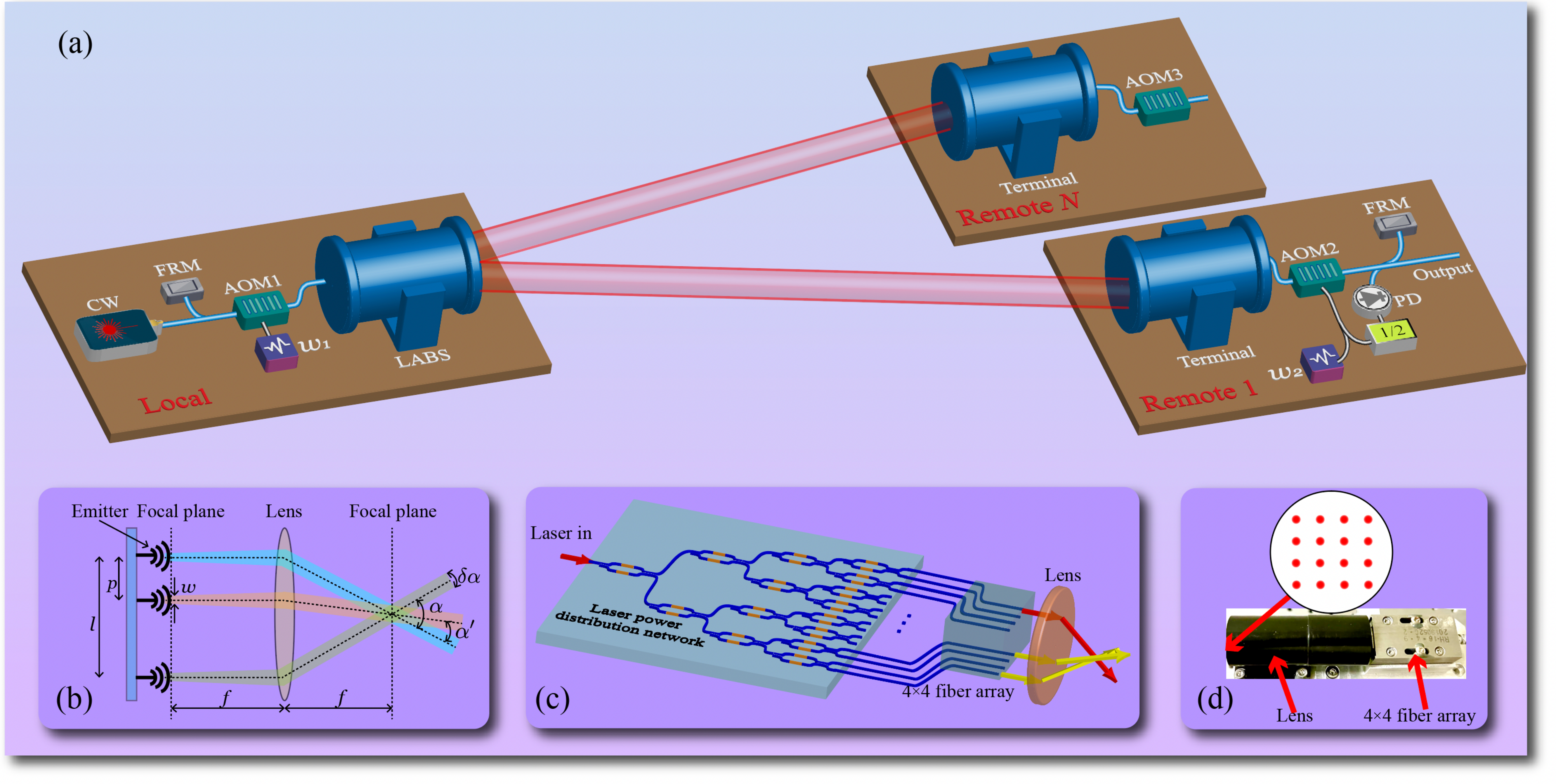}
\caption{(a) Architecture of point-to-multiplepoint optical frequency transfer with the lens assisted silica integrated beam steering. The phase noise introduced by the open air is effectively compensated by using the passive phase noise cancellation technique at each remote site as \cite{hu2020multi, xue2021branching}. (b) The principle of the lens assisted beam-steering (LABS). {Here $l$, $w$, $p$ and $f$ are {the length the of emitter array}, the size of the emitter array,  the distance between two adjacent emitters and the focal length of the lens, respectively. }(c) The power distribution networks of the LABS based on an integrated SiO$_2$ $1\times 16$ optical switch can be configured as the single output (red arrow) or arbitrary number of outputs (i.e., two outputs as shown in  the yellow arrows) with different beam angles. (d) The photograph of the packaged fiber array and lens. {Upper figure shows the distribution map of 16 beams output of the LABS.} FRM: Faraday mirror, PD: photo-detector, CW: continue-wave optical reference, AOM: acousto-optic modulator.}
\label{fig1}
\end{figure*}

Here we demonstrate a point-to-multiplepoint optical frequency transfer platform based on the LABS. The LABS, shown in Fig. \ref{fig1}, is based on a lens that deflects the light to any number of outputs with different far-field angles determined by the integrated switch network. The demonstrated LABS has 16 scanning angles and can support both single- and multiple-user at different locations via generating multiple beams by properly setting the switch network. At the same time, each beam's power can be flexibly distributed  to achieve a balance of  the distance (loss) and the number of users. A proof-of-concept $1\times2$ point-to-multiplepoint optical frequency transfer is demonstrated. Experimental results demonstrate that high performance of optical frequency transfer can be achieved and two users at different locations with the free-space links of 24 m and 50 m can work simultaneously. Each of them can support optical frequency transfer with a fractional instability of better than $4.5\times10^{-17}$ at the integration of  1 s and $7.7\times 10^{-20}$ at 20,000 s. This demonstration is the first lens-assisted  beam steering application for  optical frequency transfer in open air and shows the great potential for point-to-multiplepoint optical frequency transfer in the chip-scale monolithic transceivers. This proposed technique proves the two-dimensional LABS in the point-to-multiplepoint optical frequency transfer applications and also provides a promising way for developing a fully integrated point-to-multiplepoint optical frequency transfer system.

\section{LABS assisted point-to-multiplepoint optical frequency transfer}
\subsection{Principle}

The architecture of the proposed point-to-multiplepoint optical frequency transfer system based on the use of the LABS is illustrated in Fig. \ref{fig1}(a). {The principle of stabilized point-to-multiplepoint optical frequency transfer is similar with our previous fiber-based point-to-multiplepoint optical frequency transfer  \cite{hu2020multi, xue2021branching}. The main idea is to obtain the beatnote by heterodyne beating the single-pass light against the triple-pass light onto a photo-detector (PD) at the remote sites, which acts as a phase detector for extracting the phase noise introduced by each link. The extracted phase noise is fed into the remote acousto-optic modulator (AOM) to compensate the phase noise. With this open-loop design, namely PNC, neither active phase discrimination nor dynamic phase compensation is required, enabling to get rid of the servo bump appeared in the active phase noise cancellation technique \cite{hu2020multi, xue2021branching}.}

One main difference from the fiber-based point-to-multiplepoint optical frequency transfer  is that here we use the LABS to replace the fiber splitter in \cite{hu2020multi, xue2021branching}. The centralized architecture serving as the local site is used to send an ultrastable optical reference, where the optical reference is steered by the LABS based on the users' locations to provide free-space connectivity. At the same time, we configure the switch network, {which} depends on the position and number of the users. The LABS is designed to operate under a coaxial optical frequency transfer {configuration} at 1550 nm. The coaxial design sharing the same transmitting and {receiving} light paths enables the reduction of the phase noise coming from the out-of-loop path \cite{xue2021branching}. The light coming from the optical reference is injected into the point-to-multiplepoint optical frequency transfer system. {To distinguish the undesired light  from the reflected light at the remote sites, we still install AOMs at the both sites. At the same time, the remote AOM is also used for both compensating the phase noise introduced by the air path. To label unique frequency for each remote site, we have the different driving radio frequency (RF) for each remote AOM. }

{The detailed working procedure is as follows. After retro-reflecting the optical reference signal between the local site and the remote site, taken the remote site 1 as an example, the optical signal output from the remote site 1 is,
\begin{equation}
\begin{split}
E_1&\propto\sum_{m=1}^{\infty}\cos\left[(\nu+(2m-1)(\omega_{1}-\omega_{2}))t+\phi_s\right.\\
&\,\,\,\,\,\,\,\,\,\,\,\,\,\,\,\,\,\,\,\,\,\,\,\,\,\,\,\,\,\,\,\,\,\,\,\,\,\,\left.+(2m-1)(\phi_p+\phi_1-\phi_2)\right],
\end{split}
\label{eq7}
\end{equation}
where  $m$ delegates the number of the trip times propagating in the free link connecting the local site and the remote site 1, $\nu$ and $\phi_s$ denote the optical reference's angular frequency and phase, $\omega_1$ ($\omega_2$) and $\phi_1$ ($\phi_2$) represent  the RF signal's  angular frequency and  phase for the local (remote) AOM, and $\phi_p$ is the phase noise introduced by the single-pass free-space channel. }

At the remote site 1, the beatnote recovered by a photo-detector (PD) by heterodyne beating the single-pass ($m=1$) light against the multiple-pass ($m\geq2$) light can have a form of, 
 \begin{equation}
E_{2}\propto\sum_{i=1}^{\infty}\cos[2i(\omega_{1}-\omega_{2})t+2i(\phi_p+\phi_1-\phi_2)].
\end{equation}

To extract the phase noise coming from the free-space channel, we select an RF signal when $i=1$ by using an RF bandpass filter, and divide the angular frequency of the selected RF signal  with a factor of 2, producing,
\begin{equation}
E_{3}\propto\cos((\omega_{1}-\omega_{2})t+(\phi_p+\phi_1-\phi_2)).
\end{equation}

Afterwards, we feed $E_{3}$ together with $E_r=\cos(\omega_2 t+\phi_2)$ into the remote AOM2 with an assistance of the RF power combiner. Afterwards, the output signal from the remote site can be expressed as,
\begin{equation}
\begin{split}
E_4&\propto\cos(\nu t+\phi_s)+\sum_{m=2}^{\infty}\cos\left[(\nu+(2m-1)(\omega_{1}-\omega_{2}))t\right.\\
&\,\,\,\,\,\,\,\,\,\,\,\,\,\,\,\,\,\,\,\,\,\,\,\,\,\,\,\,\,\,\,\,\,\,\,\,\,\,\left.+\phi_s+(2m-1)(\phi_p+\phi_1-\phi_2)\right].
\end{split}
\label{eq7}
\end{equation}

Note that the first term shown in  Eq. \ref{eq7} is simultaneously not dependent on the phase noise coming  from the free-space channel and the phase noise introduced by the  RF references at both sites.  Importantly, the characteristic of the fast phase recovery speed owned by the adopted passive phase noise cancellation technique \cite{hu2020cancelling, hu2020multi, xue2021branching} is beneficial for interruptions frequently occurred optical frequency transfer over free-space, due to the strong atmospheric turbulence \cite{giorgetta2013optical, kang2019free}.

\subsection{LABS device}
In point-to-multiplepoint optical frequency transfer  based on the LABS, {the core characteristics are }the LABS's the beam-steering coverage area, and the number of accessible users. Our LABS is similar with the one used for the Lidar application as described in \cite{cao2020lidar, li2019lens}. In this section, we will briefly describe the  design and measured results of the LABS used for our point-to-multiplepoint optical frequency transfer application.  The general principle of the LABS is shown in Fig. \ref{fig1}(b). The beams generated from the emitters are parallel. Here the emitters array  is at the front focal plane of the lens. With this configuration, the lens can be simultaneously used for collimating and steering the beams.

 \begin{figure}[htbp]
\centering
\includegraphics[width=1\linewidth]{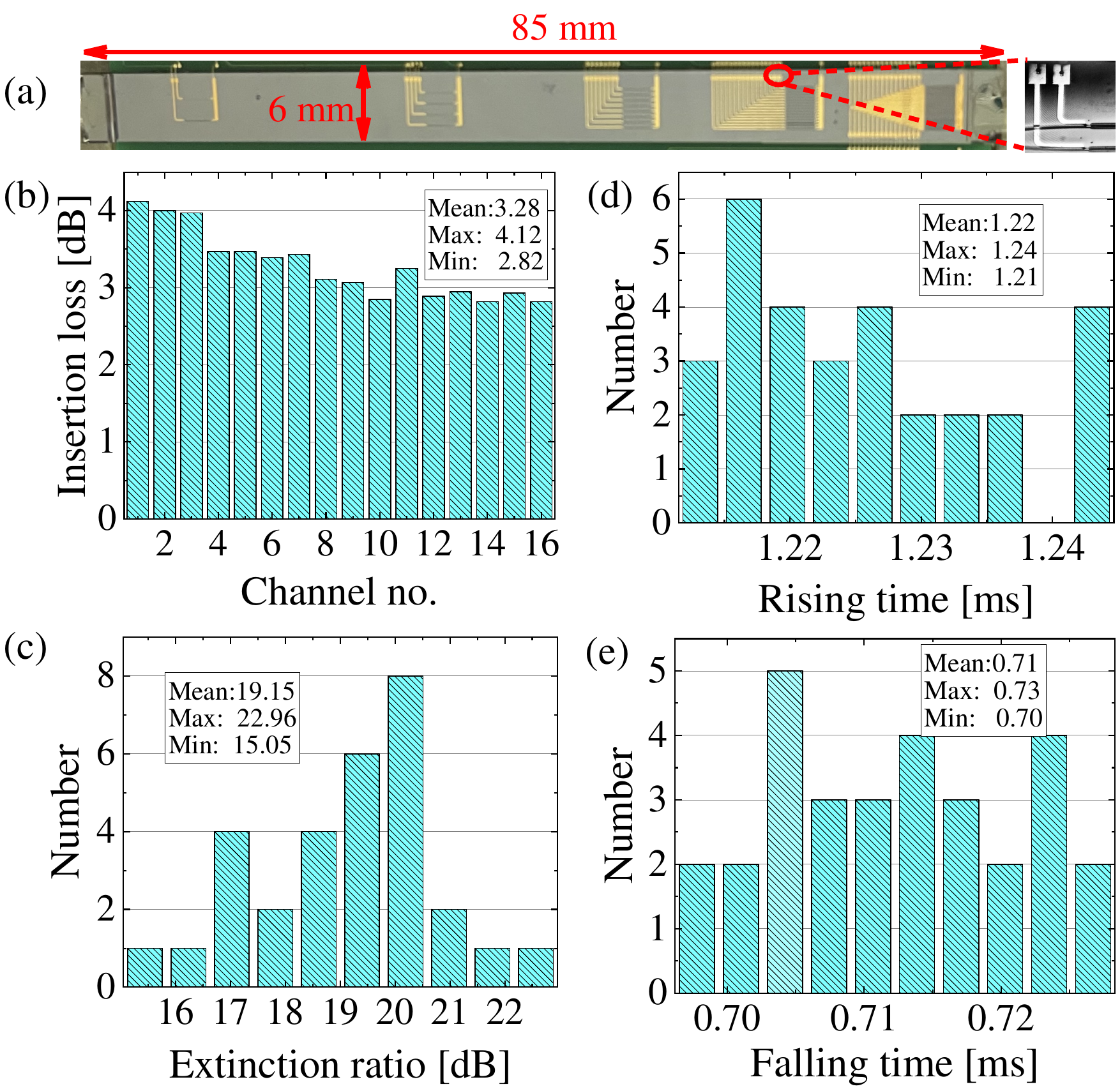}
\caption{(a) Photograph of the SiO$_2$ $1\times16$ switch chip. The inset shows the microscopic image of the part of the chip. (b) Measured insertion loss of the 16 optical channels. (c) Measured extinction ratio for the all 15 Mach-Zehnder (MZM) units' outputs. (d) and (e) Measured rising and falling switch time for all the MZM units, respectively.}
\label{fig2}
\end{figure}
 
The proposed LABS components mainly include an integrated SiO$_2$ $1\times16$ switch chip, a lens, and a $4\times4$ fiber array as illustrated in Fig. \ref{fig1}(c).  A photograph of our  SiO$_2$ chip is shown in Fig. \ref{fig2}(a). The size of the chip is $85 \times 6$ mm$^2$. {The switch is installed on the ceramic substrate, which equips a cooling fin for dissipating the heat.} To reduce the size of the chip, silicon photonics platforms can be used because it  can offer 100-fold higher  integration density than the conventional SiO$_2$-based switches with the  complementary metal oxide semiconductor (CMOS) fabrication processes \cite{chen2012compact, suzuki2014ultra, zhao201616}. Unfortunately, additional optical losses have limited the number of ports in silicon-based switches. {Alternatively, the CMOS compatible ultra-low loss silicon nitride waveguide will be a promising solution, in which the phase shifting efficiency is high due to its high thermo-optic coefficient and small footprint compared to silica thermo-optic switches \cite{joo2018cost}.} {Figure \ref{fig2} (b) and (c) shows histograms of insertion losses and extinction ratios. The insertion loss was averaged at 3.28 dB. The extinction ratio was calculated from the difference between the on and off state losses with a mean value of about 19 dB. The low extinction ratio is mainly due to the fabrication error of the front and rear 3-dB splitters for each Mach-Zehnder (MZM) unit far from the ideal value \cite{suzuki2015ultra}. It is important to note that the effect of the crosstalk between channels introduced by the low insertion ratio can be effectively avoided by adopting the heterodyne detection technique with different frequencies for each remote site. We also measured the optical response to the heater driving voltage by applying square wave with the rising and falling time of 10 ns. The response time was less than 1.24 msec and 0.73 msec for the rising edge and falling edge as shown in Fig. \ref{fig2}(d) and (e), which has the same level with that of a conventional SiO$_2$-based switch \cite {shibata2003silica}. There was no degradation in the performance of the optical switch after months of operation.}

To {convert} the 1D chip outputs in the 2D grid, we connect the 16 outputs of the chip to a $4\times4$ fiber array with a spacing of 0.25 mm to a 2D array. This 2D fiber array can be replaced by the 2D grating array as recently demonstrated in \cite{chang20212d}.  In order to collimate and steer the beam for long distance optical frequency transfer, here we use a lens with a focal length and diameter of 40 mm and 20 mm, respectively. With this configuration,  we can have the beam steering resolution of $\alpha'=\tan^{-1}\left(p/f\right)\simeq0.35^{\circ}$, the maximum steering angle of $\alpha=2\tan^{-1}\left({l}/{2f}\right)\simeq1.05^{\circ}\times1.05^{\circ}$ and the divergence angle of $\delta\alpha=\tan^{-1}\left({w}/{f}\right)\simeq0.014^{\circ}$. Under the present configuration, the covering area is constrained by the small spacing between fiber arrays.  It is important to stress that other beam divergence, beam-steering angles, and beam-steering resolutions, can be easily reconfigured  by properly rearranging the emitter array and lens parameters.  For example, the maximum steering angle can be increased to $26.57^{\circ}\times26.57^{\circ}$ by increasing the emitter spacing to 2.5 mm and replacing the fiber array with a $16\times16$ one.


One key advantage of the LABS based optical frequency transfer is supporting both  single-user and multiple-user applications. Arbitrary number of the users can be configured by changing the phase shifters to guide the input signal to  power splitters to generate arbitrary number of signals as the one (red) and two outputs (yellow) shown in Fig. \ref{fig1}(c). {Different power allocations can flexibly configure  the $1\times16$ switch, where larger powers are allocated to farther users.} By doing this way, optical frequency transfer to multiple users at different locations can be achieved.

\section{Experimental setup and results}
\subsection{Experimental setup}

\begin{figure}[htbp]
\centering
\includegraphics[width=1\linewidth]{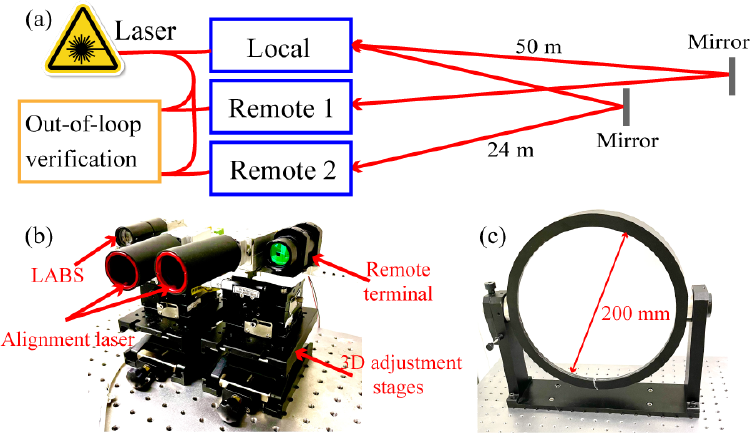}
\caption{(a) Experimental setup.  It includes one local and two remote sites. Here all the sites are collocated and the link is folded for the out-of-loop verification. (b) Photograph of the LABS-based  and remote terminals mounted on a 3D adjustment stages consisting of one vertical translation stage, one rotation stage and one tilt stage. Each terminal is equipped  with {an alignment} laser at the wavelength of 532 nm beam for the beam alignment. (c) Photograph of the intermediary flat mirror with the diameter of 200 mm. }
\label{fig3}
\end{figure}

Figure \ref{fig3}(a) shows the schematic diagram  of the experimental setup. The local and remote sites are collocated for the purpose of the out-of-loop measurements  by folding the optical free-space link via an intermediary flat mirror \cite{hu2021performance}.  To demonstrate the proposed technique, we set the experimental setup  as displayed in Fig. \ref{fig3} (a) with the two remote sites (remote site 1 and 2) due to the equipment constraints. In order to develop a folded link, a flat mirror with the diameter of 200 mm is used at the middle of the free-space link as shown in Fig. \ref{fig3}(c). The local site and one remote site are collocated and mounted on an optical table as shown in Fig. \ref{fig3}(b). To achieve the beam alignment between the local and remote sites, the  terminals are {installed on two independent 3D adjustment stages}. Each stage consists of one vertical translation stage, one rotation stage and one tilt stage. {At each remote  transceiver, the light output from optical fiber is collimated by a custom-made lens with the focal length of 75 mm and the clear aperture of 25 mm, resulting in the beam waist radius of 7.4 mm.} At the same time, each terminal equipped with an  {alignment} beam at the wavelength of 532 nm for the assistance of the beam alignment. {The alignment procedure between the local and remote site  is as follows. We first coarsely align the beam with the assignment of the alignment lasers by adjusting both 3D translation stages. Usually, after this stage, some light has successfully transmitted from the local site to the remote site. Then we continue to adjust the two 3D translation stages for optimizing the insertion loss.}    The optical reference is a narrow-linewidth optical source (NKT X15), which has the frequency near 193 THz and the output power of 13 dBm. The free-space link is located in the corridor of the laboratory on the 1st floor of Building 5 of the School of Electronic Information and Electrical Engineering (SEIEE) of Shanghai Jiao Tong University. The signal was injected into the local site based on the LABS and was then transmitted along a 24 m fiber free-space link to the remote site 1 and 50 m to the remote site 2. At each site, 30 m fiber is used to connect the light to the terminal. {To combat the slow drift of the polarization of the light, we use a manual paddle fiber polarization controller to adjust the polarization per day \cite{hu2021performance}. If the temperature fluctuations become more serious, an automatic polarization control system can be used by choosing an appropriate algorithm to control the polarization as demonstrated in \cite{hu2021performance}. } Under this scenario,  the measured loss for the both single-pass links has approximately 15 dB including  3 dB from the LABS, 4 dB from the fiber coupling and the free-space channel, 8 dB from the local and remote fiber components (AOMs, optical couplers etc.). {Due to the high insertion loss between two sites, we expect 8 remote sites can be simultaneously accommodated. Actually, the fading signal can be effectively boosted by optical amplifier such as bidirectional erbium-doped fiber amplifier (BEDFA). This has been widely adopted in the fiber-based optical frequency systems \cite{ma1994delivering, predehl2012920, Calonico2014, cantin2021accurate, hu2020cancelling}. For example, to boost single-pass light by more than 15 dB, the maximum accommodation remote sites will be increased to 16.} Here  we use $\omega_l=2\pi\times75$ MHz for the AOM1 (upshifted, $+1$ order),  $\omega_{r}=2\pi\times45$ MHz for the AOM2 (downshifted, $-1$ order) and $\omega_{r}=2\pi\times93$ MHz for the AOM3 (downshifted, $-1$ order). To match the bandwidth of the available AOMs, we mix the beatnote with an assistant RF signal of 5 MHz (101 MHz). Then the upper (lower) sideband of the mixed signals with a frequency of 35 MHz (83 MHz) for the remote site 1 (2) is used. After being compensated, we can evaluate the out-of-loop beatnote of $40$ MHz (8 MHz) for the remote site 1 (2).

\subsection{Frequency domain characterization}

\begin{figure}[htbp]
\centering
\includegraphics[width=1\linewidth]{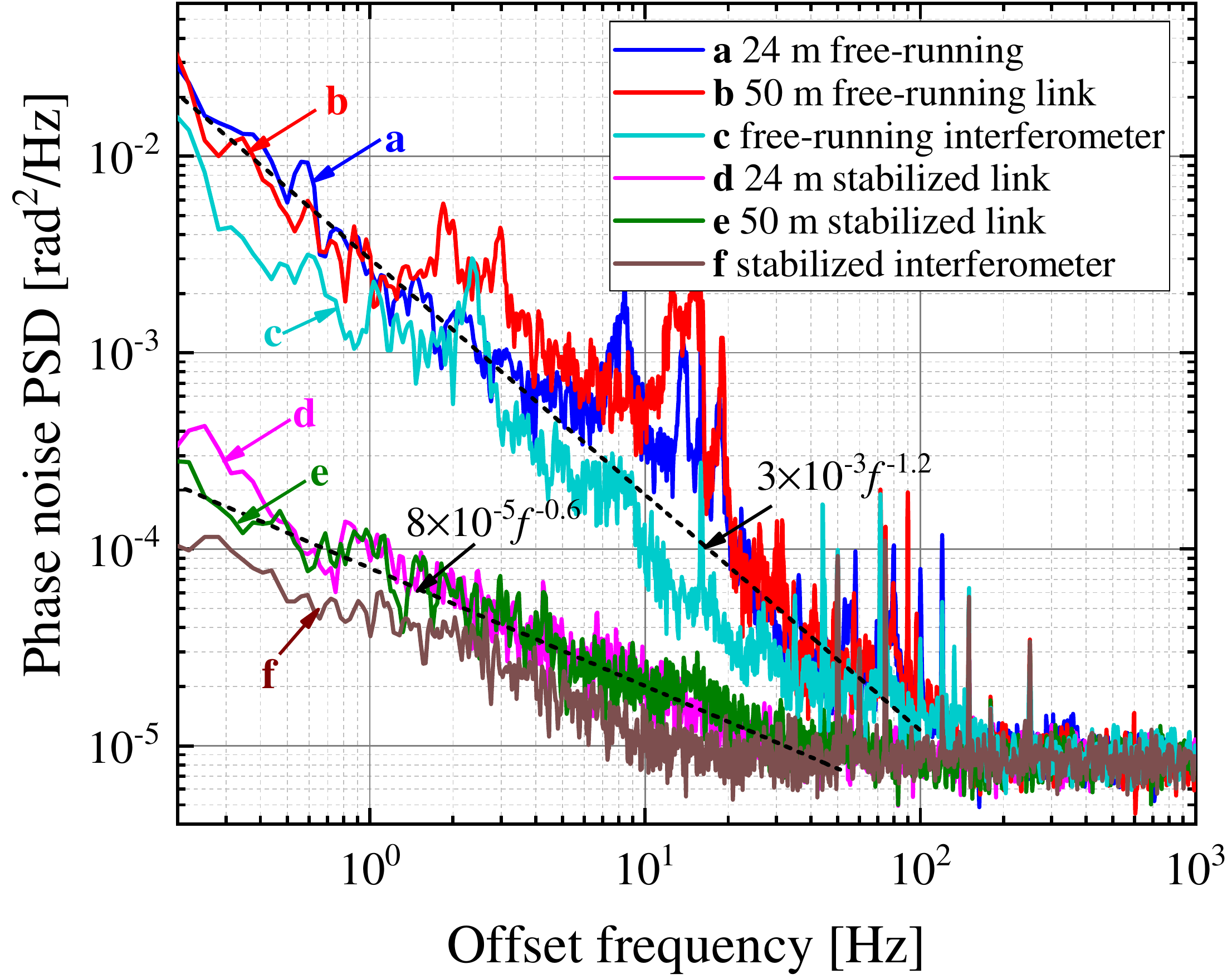}
\caption{Measured phase noise PSDs of \textbf{a} the 24 m free-running free-space link (blue curve), \textbf{b}  the 50 m free-running free-space link (red curve), \textbf{c}  the free-running fiber interferometer (cyan curve), \textbf{d}  the stabilized 24 m free-space link (magenta curve), \textbf{e}  the stabilized  50 m free-space link (olive curve), and \textbf{f}  the stabilized  fiber interferometer (wine curve).}
\label{fig4}
\end{figure}

Figure \ref{fig4} shows the phase noise power spectral densities (PSDs) of the 24 m free-space link, the 50 m free space link and the fiber interferometer with and without the implementation of the passive phase noise cancellation technique. The phase noise PSDs of the free-running 24 m link (\textbf{a}, blue curve) and the 50 m link (\textbf{b}, red curve) are shown.  Below $f<100$ Hz, the free-running curves scale down with a slope of about $3\times10^{-3}f^{-1.2}$, and reaching around $10^{-5}$  rad$^2/$Hz after 100 Hz. Their noises of the free-running 24 m link (\textbf{a}, blue curve) and the 50 m link (\textbf{b}, red curve) slightly differ because the measurements are done by two independent times. We can see that the free-running system is mainly limited by the flicker phase noise and white frequency noise sources \cite{barnes1971characterization}. The data for the unstabilized links exhibit a slope different from the theoretical Kolmogorov spectrum with the trend of $f^{-8/3}$ \cite{gozzard2018stabilized, djerroud2010coherent}, illustrating that the system is consisting of  other noise sources except for the noise from the free-space link. As a comparison, we also measured the unstabilized fiber interferometer as illustrated by the cyan curve (\textbf{c}), demonstrating that the noise of the current system is partially constrained by the noise of the unstabilized fiber interferometer. Both passive phase stabilized results are similar,  demonstrating the {slope} of $8\times10^{-5}f^{-0.6}$, scaling from $8\times10^{-5}$ rad$^2/$Hz at 1 Hz down to $10^{-5}$ rad$^2/$Hz at 30 Hz. {The noise of the stabilized link mainly includes the flicker phase and white phase noise sources. By comparing the stabilized free-space link with the stabilized fiber interferometer, we can conclude the system is mainly limited by optoelectronic devices and RF components used in the system introduced additional $1/f$ noise \cite{boudot2012phase}.} For both links, the phase noise is suppressed by approximately a factor of 40 at 1 Hz by implementing passive phase noise cancellation \cite{hu2020multi, xue2021branching}.  {Interestingly, we can clearly notice that the servo bumps are removed in our passive phase noise cancellation technique as observed in our previous work \cite{hu2020multi, xue2021branching}.}

\subsection{Frequency instability characterization}

\begin{figure}[htbp]
\centering
\includegraphics[width=1\linewidth]{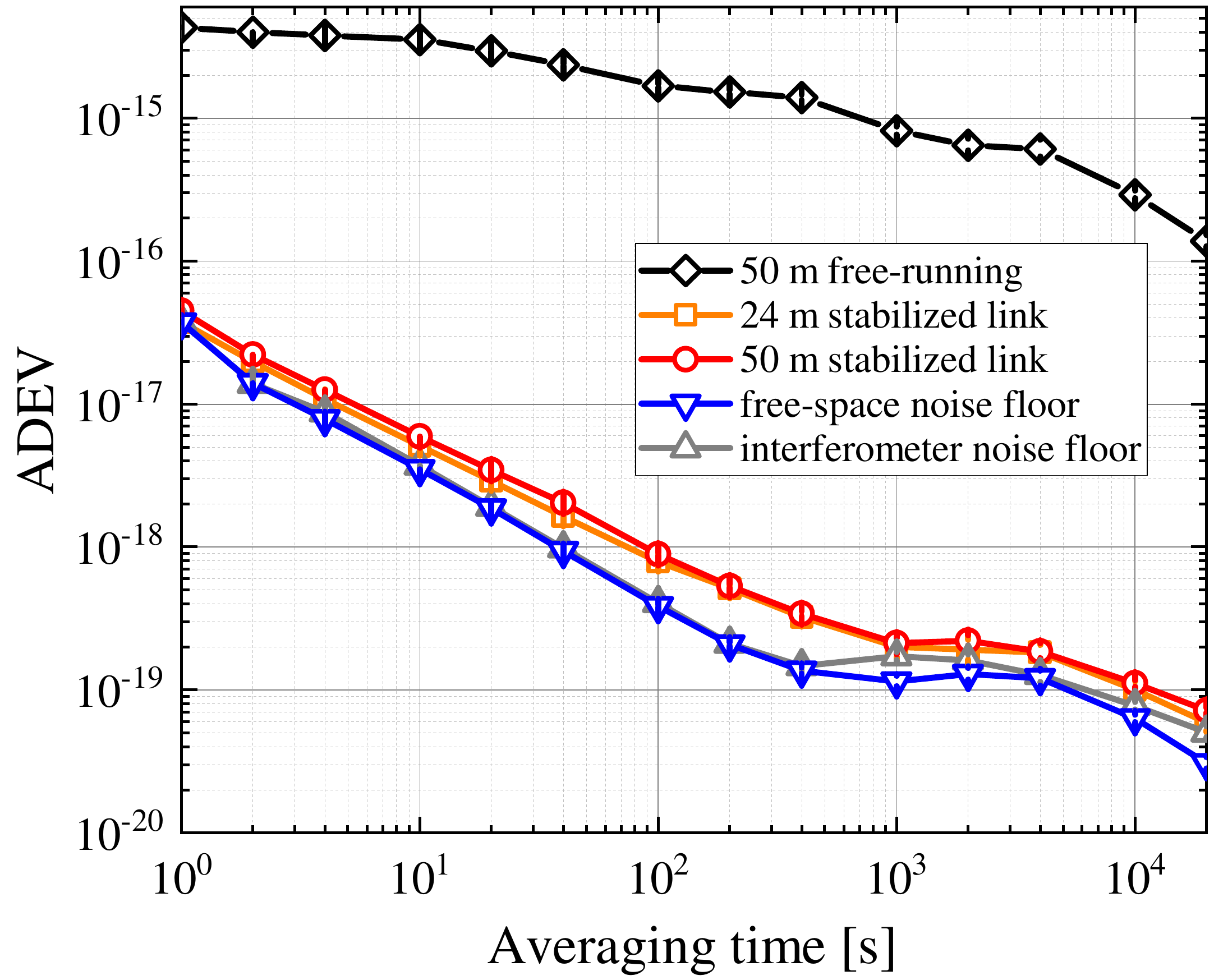}
\caption{Measured fractional frequency instability, calculated from $\Pi$-type data with the ADEV of the 50 m free-running link (black  diamonds), the stabilized 24 m link (orange squares), the compensated 50 m link (red circles), and the stabilized free-space noise floor connected by a 1 m free-space link (blue down triangles) and the fiber interferometer noise floor (gray up triangles). For the fiber interferometer noise floor, the free-space link and terminals are removed.}
\label{fig5}
\end{figure}

\renewcommand{\arraystretch}{1.5} 
\begin{table*}[ht]
	\centering
		\caption[1]{\label{tab2} Comparison of representative free-space optical time or frequency transfer systems' performance. ANC: active phase noise cancellation, PNC: passive phase noise cancellation, CW: continuous wave.}
		\begin{tabular}{ccccccc}
			\hline
			\hline
			    Method     &    Laser source     &    Distance     &     Noise floor        &    System fractional instability    &     Users     &     Reference\\
			\hline 
			   Two-way     &    Fiber comb     &    2 km     &     $1\times10^{-15}$/1 s, $1\times10^{-18}$/1,000 s      &    $1\times10^{-15}$/1 s, $1\times10^{-18}$/1,000 s     & 1 & \cite{giorgetta2013optical}   \\
			Two-way     &    Fiber comb     &    4 km     &     $0.95\times10^{-17}$/1 s, $6\times10^{-20}$/850 s      &    $1.2\times10^{-17}$/1 s, $6\times10^{-20}$/850 s     & 1 & \cite{sinclair2018comparing}   \\
			Two-way     &    Fiber comb     &    16 km     &     $3\times10^{-15}$/1 s, $3\times10^{-20}$/10,000 s      &    $3\times10^{-14}$/1 s, $4\times10^{-18}$/3,000 s     & 1 & \cite{shen2021experimental}   \\
			    Two-way     &    Fiber comb     &    28 km     &     $1\times10^{-15}$/1 s, $1\times10^{-18}$/1,000 s      &    $2\times10^{-15}$/1 s, $6\times10^{-19}$/1,000 s     & 1 & \cite{bodine2020optical}   \\
			    ANC     &    CW     &    18 km     &     /     &    $3\times10^{-15}$/0.1 s    & 1 & \cite{kang2019free}   \\
			   ANC     &    CW      &    600 m     &     $4.5\times10^{-18}$/1 s, $1.2\times10^{-18}$/64 s      &    $8.9\times10^{-18}$/1 s, $1.3\times10^{-18}$/64 s     & 1 & \cite{gozzard2018stabilized}   \\
			   ANC     &    CW      &    265 m     &     /     &    $2.5\times10^{-18}$/1 s, $1.6\times10^{-19}$/40 s      & 1 & \cite{dix2021point}   \\
			    ANC     &    CW      &    17 m     &     /     &    $1.0\times10^{-16}$/1 s, $2.3\times10^{-18}$/250 s     & 1 & \cite{vishnyakova2020optical}   \\
			    ANC     &    CW      &    2.4 km     &     /     &    $4.0\times10^{-16}$/1 s, $6.1\times10^{-21}$/300 s     & 1 & \cite{gozzard2021ultra}   \\
			   PNC     &    CW      &    50 m     &     $3.7\times10^{-17}$/1 s, $3.0\times10^{-20}$/20,000 s      &    $4.5\times10^{-17}$/1 s, $7.7\times10^{-20}$/20,000 s     & {up to 16} &This work  \\
			\hline
			\hline
		\end{tabular}
\end{table*}

\begin{figure*}[htbp]
\centering
\includegraphics[width=1\linewidth]{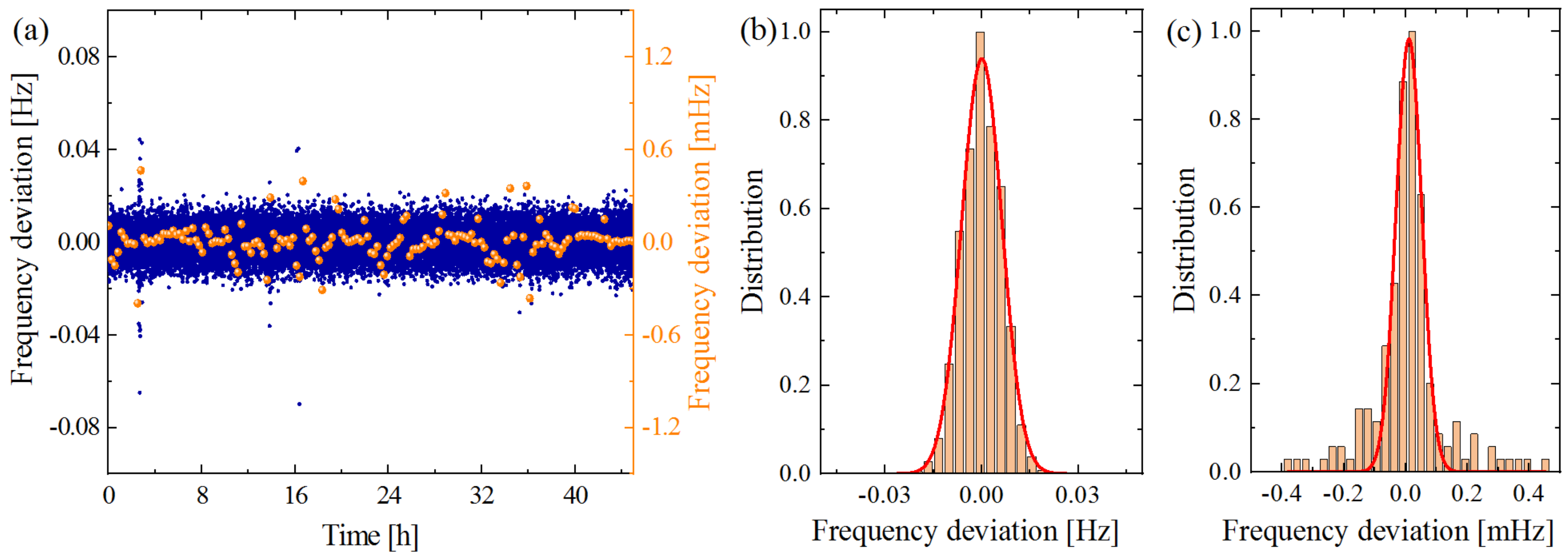}
\caption{Measured frequency beatnote between input and stabilized output signal over the 24 m free-space link. (a) 163,498 data points measured by a $\Pi$-type frequency counter at a 1 s gate time (blue points, left axis).  The unweighted mean ($\Pi$-type) values for all cycle-slip free 1,000 s average (163 data points) as the brown dots are shown in the right frequency axis. Histograms (brown bars) and Gaussian fits (red curves) for (b) 163,498 data points and (c) 163 data points.}
\label{fig6}
\end{figure*}

Figure \ref{fig5} illustrates the fractional frequency instability of both free-running and stabilized optical frequency transfer, calculated from $\Pi$-type data with an Allan deviation (ADEV).  The  fractional frequency instability of the stabilized 24 m (50 m) link is $3.7\times10^{-17}$ ($4.5\times10^{-17}$) at the 1 s averaging time, and reaches approximately $5.9\times10^{-20}$ ($7.7\times10^{-20}$) at $20,000$ s.  The fractional frequency instability results  show that the passively phase stabilized optical frequency transfer case effectively suppresses the free-space link noise by around 3 orders of magnitude. {The system performs the similar fractional frequency instability with free-space optical frequency transfer systems \cite{dix2021point, kang2019free, gozzard2018stabilized} and comb-based optical carrier-phase tracking assisted time-frequency transfer over a 4 km turbulent link \cite{sinclair2018comparing}, and exceeds more than one order of magnitude than comb-based 2 km time-frequency transfer \cite{giorgetta2013optical}} as listed in Tab. \ref{tab2}. Between 1,000 s and 4,000 s,  the fractional frequency instability values remain the same level, which is mainly attributed to the phase noise on the outside paths coming from imperfect temperature stabilization in the present set-up \cite{hu2020passive, tian2020hybrid}. {The fractional frequency variations $y$ with the outside path length of $\delta l$ with the time-varying temperature of $T(t)$ can be calculated as \cite{stefani2015tackling, tian2020hybrid},
\begin{equation}
y =\frac{\Delta\nu_0}{\nu_0}=\frac{1}{c}\left(\delta l\frac{dn}{dT}+n\frac{dl}{dT}\right)\frac{dT(t)}{dt}
\end{equation}
{where $\nu_0$ and  $\Delta\nu$ are the optical carrier frequency and its frequency shift, and $c$ denotes the speed of light.}  In our typical experimental condition with the sinusoidal temperature fluctuation amplitude of 10 mK, we can estimate the ADEV as high as $4\times10^{-20}$ when the out-of-loop is up to $\delta l=10$ cm, which is corresponding  to the experimental results.} Beyond 4,000 s, the value starts to decrease again thanks to the high performance of the two-layer temperature stabilization system with the better long-term stability \cite{hu2021performance}.  Similar fractional frequency instability trends have been reported in the previous experimental results with the similar temperature stabilization performance \cite{xu2018studying, xu2019reciprocity}. The results for the short 1 m free-space link  and fiber interferometer as  the down and up triangle markers illustrated in Fig. \ref{fig5} again confirm that the degradation of the fractional frequency instability after 1,000 s is due to the outside phase noise. We think better thermal isolation of the out-of-loop components or integrating the optical paths and components into a single photonic chip will improve the measured performance of the system between 1,000 s and 4,000 s \cite{hu2021performance, hu2021silicon}.

\subsection{Frequency uncertainty analysis}

Complementary to the fractional instability and phase noise PSDs characterizations, the uncertainty has to be carefully examined by calculating the mean value of the beatnote between the input and output signals. Figure \ref{fig6}(a) illustrates the beatnote’s data for the stabilized 24 m free-space link, taken with a 1 s gate time and $\Pi$-type counters, over successive 163,498 s as the blue points shown in the left axis.  As the yellow points shown in right axis in  Fig. \ref{fig6}(a), the arithmetic mean of all cycle-slip free 1,000 s intervals results in 163 points. A few large frequency deviations shown in Fig. \ref{fig6}(a) are mainly coming from the human activity in the corridor. Figure \ref{fig6}(b) and (c) plot the histograms and Gaussian fits of the frequency deviations for the stabilized 24 m link. Based on the results analyzed in Fig. \ref{fig6}(c), the mean frequency and the standard deviation of the 163 data points are  10.7 $\mu$Hz ($5.5\times10^{-20}$) and  41.6 $\mu$Hz ($2.2\times10^{-19}$), respectively.  Assuming that each datum for the 163 points are uncorrelated with any other (i.e. independent), the uncertainty can be calculated by $2.2\times10^{-19}/\sqrt{163}\simeq1.7\times10^{-20}$ \cite{barnes1971characterization, lee2009uncertainty, benkler2015relation}. In comparison with the long-term fractional instability of optical frequency transfer  displayed in Fig. \ref{fig5}, we can conservatively estimate the relative frequency uncertainty of $5.9\times10^{-20}$.

By using the same analysis method presented above, the mean frequency for the 50 m stabilized free-space link with the total 93,565 $\Pi$-type counter data points is -5.5 $\mu$Hz ($-2.8\times10^{-20}$). At the same time, we divide all the data into 93 groups. According to the mean value for each group,  we can obtain the standard deviation of the 93 points is 19.0 $\mu$Hz ($9.8\times10^{-20}$). Taking the fractional instability shown in Fig. \ref{fig5}, we calculate that the mean frequency offset is $-2.8\times10^{-20}$  and the statistical uncertainty of $7.7\times10^{-20}$  for the 50 m stabilized free-space link. Consequently, no systematic frequency shift within the level of a few $10^{-20}$ in the point-to-multiplepoint optical frequency transfer within the 50 m free-space link.

Overall, the system performance in terms of  the phase noise PSDs, fractional frequency instabilities and uncertainty results are all suitable for the optical frequency transfer application for the state-of-the-art optical atomic clocks \cite{grotti2018geodesy, Riehle:2017aa, delva2017test, lisdat2016clock, takamoto2020test}.

\section{Discussion}


This is the first demonstration of an LABS-based coherent point-to-multiplepoint optical frequency transfer system. In Fig. \ref{fig3}, the distance of the optical frequency  system was mainly limited by the laboratory space constraint. {In an atmosphere channel, when an optical signal is transmitted across free-space it is attenuated due to different physical effects resulting in optical loss including beam divergence, turbulence, absorption and scattering. Among them, the atmosphere turbulence is a dynamic process and strongly depends on inhomogeneous temperature and pressure introduced various refractive index values in the atmosphere \cite{bodine2020optical}.  Taking into account the beam divergence, turbulence, absorption and scattering losses in the clear weather, we can estimate the maximum transfer distance can reach up to 1.4 km in our experimental configuration when the maximum acceptable channel loss is approximately 30 dB. By increasing the beam waist radius of the local and remote terminals to 20 mm similar with \cite{bodine2020optical}, the transfer distance can be extended to 17 km, which has the same level with the comb-based time-frequency transfer \cite{bodine2020optical}. Moreover, increasing the acceptable channel loss by, such as, adopting comb-based optical time-frequency transfer \cite{shen2021experimental}, we can further extend the transfer distance. } To achieve such long optical frequency transfer distances, active optical terminals will be necessary to adjust the pointing error \cite{giorgetta2013optical, dix2021point, kang2019free, bergeron2019femtosecond, gozzard2018stabilized, sinclair2018comparing, deschenes2016synchronization}. We are currently planning to combine the LABS with active optics, such as, by using microelectromechanical (MEMS) actuated suspended silicon photonic waveguide gratings \cite{errando2019low, cao2020lidar}, which can achieve 2D beam alignment. Besides the beam pointing error correction, we expect the long distance optical frequency transfer will be limited by the power loss due to the atmosphere turbulence. 

{Importantly,  except for optical loss introduced by the free-space channel, the ultimate limitation of the system performance affected by the effects of atmospheric phase noise should be carefully studied. Although we only demonstrated  LABS based optical frequency transfer indoor, the system could be used for  free-space optical frequency transfer outdoor as the system is inherently insensitive to ambient light as the coherent detection method used. To better understand the impact of atmospheric phase noise on the propagation of light through the atmosphere, we estimate the atmospheric phase noise by using the Kolmogorov spectrum model, which can have a form of,
\begin{equation}
S_{\phi}(f)\simeq0.016k^2C_n^2 LV^{5/3}f^{-8/3} \,\,\,\,[\text{rad}^2/\text{Hz}]
\end{equation}
where $k$ is the wave number, $C_n^2$ is the turbulence structure constant, $f$ is the Fourier frequency, $L$ is the total path length, and  $V$ is a constant wind speed. For the homogenous phase noise along the free space link, the residual phase noise PSD after passive phase noise compensation can be calculated as \cite{hu2020multi}, 
\begin{equation}
S_r(f)=\frac{7}{3}(2\pi f\tau_0)^2S_{\phi}(f)\,\,\,\,[\text{rad}^2/\text{Hz}]
\end{equation}
where $\tau_0$ is the single-pass propagation delay. The ADEV can be expressed as \cite{barnes1971characterization, lee2009uncertainty, benkler2015relation},
\begin{equation}
\sigma_y(\tau)=\sqrt{2\int_0^{\infty}\frac{f^2}{\nu_0^2}S_r(f)\frac{\sin^4(\pi f\tau)}{(\pi f\tau)^2}}
\end{equation}
}

{In the case of $C_n^2=10^{-14}$ m$^{-2/3}$, and  $V=1$ m/s, the unstabilized 1.4 km free space link will limit the ADEV at the level of $4\times 10^{-14}$. After being compensated, the ADEV reaches $\sim2.7\times 10^{-18}\tau^{-1}$ , which is lower than our system's noise floor and suppresses the noise by more than four orders of magnitude. Under this circumstance, the system is limited by the noise floor even after 1.4 km transfer. More aggressively, transferring the optical signal over 17 km with the beam waist radius of 20 mm after being compensated, the ADEV of $1.1\times10^{-16}\tau^{-1}$ could be obtained.}




We have to stress that although we only demonstrated the LABS for the optical frequency transfer applications, the same device can also be used for, such as, comb-based two-way time-frequency transfer \cite{giorgetta2013optical}. While the demonstration is dedicated on the development of point-to-multiplepoint transfer between static nodes, it is worth to note that the chip-scale of the demonstrated system may prove useful for future satellite-based  optical time-frequency links. Unfortunately, other challenges, which have not been studied within the experiment described above, have to be solved before transferring the system into satellite-based applications. For example, the optical frequency transfer system will be affected by large Doppler shifts situation, which can not be cancelled by the phase stabilization technique \cite{bergeron2019femtosecond}.

\section{Conclusion}
\label{sec6}
We have proposed and experimentally demonstrated a novel point-to-multiplepoint optical frequency transfer system employing a lens assisted integrated photonic circuit. A $1\times16$ SiO$_2$ integrated optical switch as an integrated lens assisted integrated beam steering is used to configure any number of outputs for producing multiple beams at different angles, resulting in supporting both single-user and multi-user optical frequency transfer. A point-to-multiplepoint optical frequency transfer system with two remote sites over 24 m and 50 m free space links has been demonstrated. Measured results have shown that the achievable instability of  the transmission distance up to 50 m with the integrated photonic circuit is better than $4.5\times10^{-17}$ at 1 s and $7.7\times 10^{-20}$ at the integration time of 20,000 s. At the same time, the systematic frequency shift for the proposed system is within the level of a few $10^{-20}$.

A point-to-multiplepoint free-space optical frequency transfer network with a monolithic integrated transceiver is an important step toward the herald of the new generation of multiple-user time-frequency transfer for developing  free-space optical frequency transfer  networks. The star  topology network used here would have the  capability to satisfy any number of the user nodes. It can also be used {for} comparing two optical references at the two remote nodes without the direct line-of-sight between users. It is important to notice that the optical frequency transfer performance of the network in terms of fractional frequency  instability and uncertainty was not affected by adding more user nodes. The proposed technique opens the possibility of extending the optical frequency transfer capacity of future large-scale optical clock networks with the integrated lens assisted integrated beam steering.

In conclusion, the results presented here are the first demonstration of  optical frequency transfer over a point-to-multiplepoint free-space network. This demonstration with the integrated lens assisted  beam steering for  free-space optical frequency transfer shows the potential for multiplepoint optical frequency transfer in chip-scale monolithic transceivers. These advances show that optical frequency transfer  can be successfully performed across multiplepoint free-space optical frequency. The successful operation of this point-to-multiplepoint network therefore represents an important step toward future optical frequency  transfer networks that include chip-scale free-space optical frequency transfer devices.


\ifCLASSOPTIONcaptionsoff
  \newpage
\fi

\end{document}